\documentclass[aps,prl,twocolumn]{revtex4}

\usepackage{amssymb}
\usepackage{graphicx}
\usepackage{amsmath}
\usepackage{color}
\usepackage{units}

\begin{document}

\title{Interplay between conducting and magnetic systems in the antiferromagnetic
organic superconductor $\kappa$-(BETS)$_2$FeBr$_4$ \thanks{The work was supported
by the German Research Foundation (DFG) via the grant KA 1652/4-1.}}


\author{Mark V. Kartsovnik$^{1}\footnote{Email:mark.kartsovnik@wmi.badw.de}$}
\author{Michael Kunz$^{1,2}$}
\author{Ludwig Schaidhammer$^{1,2}$}
\author{Florian Kollmannsberger$^{1,2}$}
\author{Werner Biberacher$^{1}$}
\author{Natalia D. Kushch$^{3}$}
\author{Akira Miyazaki$^{4}$}
\author{Hideki Fujiwara$^{5}$}


\affiliation{$^{1}$Walther-Mei{\ss}ner-Institut, 85748 Garching, Germany}              
\affiliation{$^{2}$Technische Universit\"{a}t M\"{u}nchen, 85748 Garching, Germany}
\affiliation{$^{3}$Institute of Problems of Chemical Physics, 142432 Chernogolovka, Russia}
\affiliation{$^{4}$University of Toyama, 9308555 Toyama, Japan}
\affiliation{$^{5}$Osaka Prefecture University, 5998531 Osaka, Japan}


\begin{abstract}
The mutual influence of the conduction electron system provided by organic donor layers and magnetic system localized
in insulating layers of the molecular charge transfer salt $\kappa$-(BETS)$_2$FeBr$_4$ has been studied. It is
demonstrated that besides the high-field re-entrant superconducting state, the interaction between the two systems
plays important role for the low-field superconductivity. The coupling of normal-state charge carriers to the magnetic
system is reflected in magnetic quantum oscillations and can be evaluated based on the angle-dependent beating behaviour
of the oscillations. On the other hand, the conduction electrons have their impact on the magnetic system, which is
revealed through the pressure-induced changes of the magnetic phase diagram of the material.

\end{abstract}

\maketitle
\section{Introduction}
\label{intro}
Thanks to high crystal quality, relatively simple conduction band structures, and very good tunability between
various electronic states, organic charge transfer salts can often serve as model systems for studying
correlated electron physics, see, e.g., \cite{lebe08} for a review.
In particular, the family of bis\-(ethylenedithio)\-tetra\-selena\-fulvalene (BETS) salts containing transition-metal ions
like Fe3+, Mn2+, Cu2+, etc. represents perfect natural nanostructures with alternating single-molecular conducting
and magnetic layers. The charge transport in such compounds is provided by delocalized $\pi$-electrons of fractionally
charged BETS donors, whereas magnetic properties are dominated by localized $d$-electron spins in the insulating
anionic layers. While the latter usually undergo antiferromagnetic (AF) ordering at low temperatures, the ground state
of the $\pi$-electron system is determined by a subtle balance between different instabilities of the normal metallic
state and is particularly sensitive to the interlayer $\pi$-$d$ exchange interaction \cite{kobayashi_organic_2004}.

In the best-studied member of this family, $\lambda$-(BETS)$_2$\-FeCl$_4$, the low-temperature metal-insulator transition is
triggered by the AF ordering in the anionic layers \cite{brossard_interplay_1998}. Moreover, at high magnetic fields
where the AF insulating state is suppressed \cite{brossard_interplay_1998}, the $\pi$-$d$ interaction leads, via
the Jaccarino-Peter compensation effect \cite{jaccarino_ultra-high-field_1962} to a
fascinating field-induced superconducting (SC) state \cite{uji_magnetic-field-induced_2001}.
In the $\kappa$-(BETS)$_2$\-FeX$_4$ (M = Cl, Br) salts, differing from the former by the structure of the BETS layers,
the $\pi$-$d$ interaction is weaker. As a result, they preserve metallic properties and even become SC below the N\'{e}el
temperature \cite{otsuka_organic_2001}. However, the magnetic interactions remain very important, as seen, for instance,
in the high-field re-entrant superconductivity in the X = Br salt \cite{fujiwara_indication_2002,konoike_magnetic-field-induced_2004}.

In this paper we show a few examples illustrating the entanglement of the magnetic and conducting systems in
$\kappa$-(BETS)$_2$\-FeBr$_4$. In Sec.\,\ref{PD} we present data on the low-temperature phase diagram at ambient pressure
as well as under high quasi-hydrostatic pressure. In addition to already known effects, we report on some new results
revealing the influence of the magnetic system on the ground state of conduction electrons and vice versa. In
Sec.\,\ref{MQO} magnetoresistance quantum oscillations are presented. Here we focus on the effect of the Zeeman splitting
on the oscillation behaviour, which provides quantitative information on the strength of the $\pi$-$d$ coupling, but also
shows some unexpected features.

\section{Experimental}
\label{exp}
Crystals of $\kappa$-(BETS)$_2$\-FeBr$_4$ were grown electrochemically \cite{kobayashi_new_1996,fujiwara_novel_2001}
and had the form of rhombic platelets of submillimiter size and the largest surface parallel to the conducting BETS layers,
that is the crystallographic $ac$-plane. The interlayer ($\|$ $b$-axis) resistance was measured by standard 4-terminal
a.c. technique using a low-frequency lock-in amplifier. High-pressure measurements were done in a Be-Cu piston-cylinder
clamp cell with a silicon-organic liquid GKZh as pressure medium. The room- and low-$T$ pressure values were evaluated
from the resistance of a calibrated manganin coil. Magnetoresistance measurements were done at ambient pressure in magnetic
fields up to 14\,T. The samples were mounted on a holder allowing in-situ rotation of the sample around an axis perpendicular
to the external field direction. The orientation of the crystal was defined by a polar angle $\theta$ between the field and
the crystallographic $b$-axis (normal to conducting layers).

\section{Phase diagram}
\label{PD}

\subsection{High-field re-entrant superconductivity}
\label{SC}
\begin{figure}[tb]
	\centering
		\includegraphics[width=0.45\textwidth]{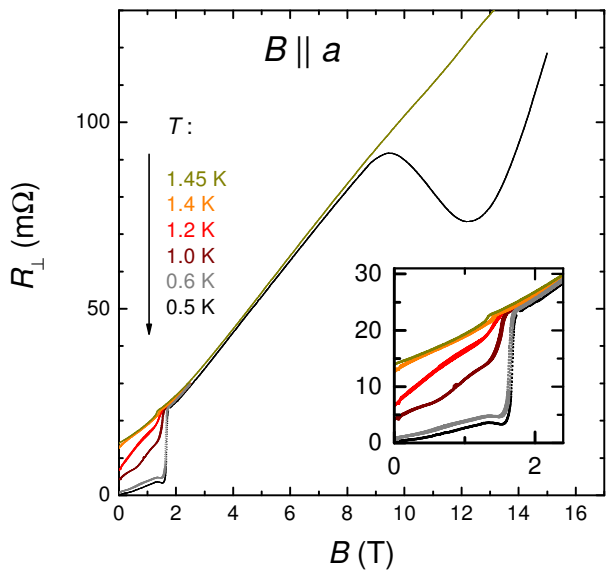}
	\caption{Interlayer resistance $R_\perp(T)$ of $\kappa$-(BETS)$_2$FeBr$_4$ as a function of magnetic field applied parallel to
the inplane $a$-axis, which is the easy magnetization axis in this system. Low-field (below 2 T) and high-field ($B > 9.5$\,T) SC
regions are clearly seen. The inset shows the low-field region in a larger scale}
	\label{RBa}
\end{figure}

The coupling between the magnetic and conduction systems is most evidently
manifested in the SC properties of $\kappa$--(BETS)$_2$FeBr$_4$. In Fig.\,\ref{RBa} we
plot examples of interlayer resistance traces in a magnetic field aligned in the conducting
layers, parallel to the crystallographic $a$-axis. This direction coincides with the easy
magnetization axis of the $d$-electron spins \cite{otsuka_organic_2001}.
Two SC regions are clearly seen, one below $\sim 2$\,T and the other at high fields, $\geq 9.5$\,T
at the lowest temperature. The latter is known \cite{fujiwara_indication_2002} as field-induced or,
here maybe better, re-entrant superconductivity. While in the present data the resistance does not
completely vanish, a full re-entrant SC transition has been observed at lower temperatures, $T < 0.3$\,K
\cite{konoike_magnetic-field-induced_2004}. From the position of the center of the SC dip one can estimate
the effective exchange field imposed on the conduction $\pi$-electron spins by the localized $d$-electron
spins aligned in the field direction, $B_{J} \approx - 12.6$\,T (the "$-$" sign indicates the AF
interaction), in perfect agreement with the earlier reports
\cite{fujiwara_indication_2002,konoike_magnetic-field-induced_2004}. This corresponds to the exchange coupling
energy $J_{\pi d}=\mu_{\mathrm{B}}gB_J/S_d \approx 0.9$\,meV, where we plugged in the Fe$^{3+}$ electron spin
$S_d=5/2$ \cite{fujiwara_novel_2001} and assumed the $g$-factor equal to 2.0 ($\mu_{\mathrm{B}}$ is the Bohr magneton).
This value is quite small; nevertheless, as we see, this interaction has a strong impact on the electronic state.

\subsection{Low-field superconductivity protected by AF ordering}
The low-field SC transition is also affected by the $\pi$-$d$ interaction. As the applied field
increases from zero, the resistance gradually increases, indicating that the material is in the resistive
flux-flow regime. However, at a certain field a sharp jump to the normal-state resistance value is observed.
Most clearly it is seen in the 0.5 and 0.6\,K curves in Fig.\,\ref{RBa}. This jump has been associated with
a transition of the Fe spin system from the AF to a paramagnetic (PM) state \cite{fujiwara_indication_2002}.
Since the external field $\mathbf{B}$ is applied along the easy axis, the Fe spins are aligned alternately
parallel and antiparallel to $\mathbf{B}$. Thus, the average exchange field on the $\pi$-electron spins is zero.
The total effective field $B_{\mathrm{eff}} = B + B_J$ is simply equal to the applied field, which is not
strong enough to kill superconductivity. As soon, as the AF order is broken, the majority of Fe spins turns
parallel to $\mathbf{B}$, creating a strong exchange field $B_j \sim -10$\,T. Therefore the effective field on
$\pi$ spins sharply increases, causing a destruction of superconductivity by the Pauli paramagnetic pair-breaking
mechanism.

The described scenario explains simultaneous suppression of the AF order in the magnetic system and of
the SC order in the conducting system. However, Konoike et al. have reported resistive
\cite{konoike_magnetic-field-induced_2004} and magnetic torque \cite{konoike_magnetic_2005} data suggesting
that the two transitions, while being close to each other, do not exactly coincide, at least at the lowest
temperatures. A possible explanation is that at low temperatures the AF state undergoes a spin-flop transition
before being completely destroyed by magnetic field. If so, the Fe spins acquire a finite average component in
the field direction. If this component is strong enough, it creates an exchange field sufficient to suppress
superconductivity in the BETS layers.

We have carefully inspected the resistive behaviour at $\mathbf{B} \| \mathbf{a}$ and found only one single
transition down to $T = 0.5$\,K, as shown in the inset in Fig.\,\ref{RBa}. Thus, if the transition splits
in two, it happens at lower temperatures.
\begin{figure}[tb]
	\centering
		\includegraphics[width=0.45\textwidth]{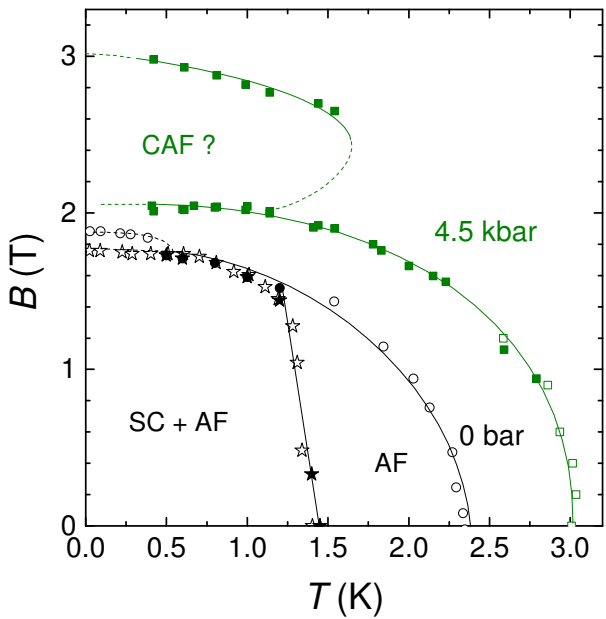}
	\caption{Phase diagram of $\kappa$-(BETS)$_2$FeBr$_4$ in magnetic field parallel to the easy axis. Circles
and stars correspond to the AF and SC transitions, respectively, recorded at zero pressure. The hollow symbols
have been plotted according to Konoike et al. \cite{konoike_magnetic-field-induced_2004}. Squares show the
transition points taken from temperature (hollow symbols) and field (solid symbols) sweeps at $p=4.5$\,kbar.
The lines are guides to the eye}
	\label{PDa}
\end{figure}
In Fig.\,\ref{PDa} we summarize the data on the $T$-dependent AF and (low-field) SC critical fields from
Konoike et al. \cite{konoike_magnetic-field-induced_2004} (hollow symbols) along with those determined in this work
(solid symbols); we now consider the zero-pressure data. One can see that the very steep, $\approx 7$\,T/K, initial slope
of the SC phase boundary near $T_c(B=0)$ is interrupted exactly when it meets the AF phase boundary. Between $1.3$
and $0.5$\,K both phase lines coincide. At lower $T$ the AF critical field seems to increase slightly above the SC one.
As noted above, the narrow region between the critical fields may be a canted AF (CAF) phase. However, more thorough
experiments are necessary to verify this.

\subsection{Pressure effect}
Fig.\,\ref{PDa} also shows the phase diagram obtained under hydrostatic pressure, $p=4.5$\,kbar. Superconductivity is
rapidly suppressed by pressure, which is usual for organic superconductors \cite{toyo07}. According to Otsuka et al.
\cite{otsuka_pressure_2004}, the critical pressure for superconductivity is 4\,kbar. In our measurements
already at 2\,kbar no trace of a SC transition has been found down to 0.42\,K.

By contrast to superconductivity, the AF ordering is enhanced under pressure. The zero-field N\'{e}el temperature
increases by $25\%$ from the ambient pressure value, qualitatively in agreement with the earlier report
\cite{otsuka_pressure_2004}. This is obviously caused by an enhancement of the exchange interactions
due to compression of the crystal lattice.

In addition to the kink feature around 2\,T, a clear second feature has been detected in the $B$-dependent
resistance at $T\leq 1.5$\,K. The positions of both kinks are plotted in Fig.\,\ref{PDa} (squares).
The behaviour resembles that observed on the isostructural sister compound
$\kappa$--\-(BETS)$_2$\-FeCl$_4$ ($\kappa$-Cl salt) at ambient pressure \cite{kunz16}. In the $\kappa$-Cl salt
this behavior was associated with a spin-flop transition into an intermediate, CAF state. Therefore, we
suggest that the same happens in the present $\kappa$--\-(BETS)$_2$\-FeBr$_4$ salt under pressure.

Comparing to the present compound, in $\kappa$-Cl the CAF phase occupies a significant part of
the ambient-pressure phase diagram \cite{kunz16}.
The difference stems from different relative contributions of the direct $d$-$d$ exchange and
indirect RKKY exchange, involving the $\pi$-$d$ and $\pi$-$\pi$ coupling, in setting the AF order.
In $\kappa$--(BETS)$_2$FeBr$_4$ the ordering is predicted to be strongly dominated by the
direct exchange, whereas in the $\kappa$-Cl salt the indirect interactions play a considerable role
\cite{mori_estimation_2002}.
The delocalized $\pi$-electrons provide much more isotropic interactions in the plane of conducting layers
than the $d$-electrons localized in the insulating layers. Therefore the magnetic anisotropy
of the $\kappa$-Cl salt is weaker, favouring a spin-flop from the easy $a$-axis to the next-easy
inplane $c$-axis at $\mathbf{B} \| \mathbf{a}$. Turning to our salt, we can speculate
that high pressure leads to a stronger enhancement of the indirect exchange, thus facilitating the
spin flop transition and making the CAF phase more stable. This looks reasonable, since the conduction
electrons in organic metals are known to be extremely sensitive to pressure \cite{toyo07}.

\section{Magnetic quantum oscillations}
\label{MQO}
\begin{figure}[tb]
	\centering
		\includegraphics[width=0.45\textwidth]{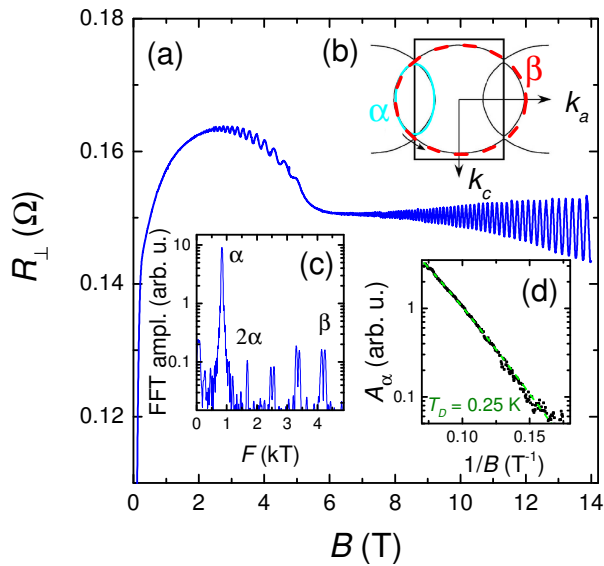}
	\caption{(a) Field-dependent interlayer resistance at $B \perp $ layers, $T= 0.45$\,K. (b) Two-dimensional Fermi
surface consisting of a closed part on the border of the Brillouin zone and a pair of open sheets along $\mathbf{k}_a$,
according to Kobayashi et al. \cite{kobayashi_new_1996}. The classical ($\alpha$) and magnetic-breakdown ($\beta$) orbits
are shown. (c) FFT spectrum of the SdH oscillations in the field window 9 to 14\,T. (d) Amplitude of the
$\alpha$-oscillations plotted against inverse field. The dashed line is the best fit using the standard theory
\cite{ShoenbergBook1984}, yielding the Dingle temperature $T_D=0.25$\,K}
	\label{SdH}
\end{figure}

The influence of the magnetic system on the normal-state charge carriers can be probed by magnetic
quantum oscillations. Fig.\,\ref{SdH} shows an example of oscillating resistance (Shubnikov-de Haas, SdH
oscillations) in a field perpendicular to conducting layers, $\mathbf{B} \| \mathbf{b}$.
The AF ordering is reflected in the behaviour of the oscillations as well as in the
nonoscillating magnetoresistance component. In Fig.\,\ref{SdH} one can distinguish two different
magnetoresistance regimes corresponding to the AF ($B\leq 5$\,T) and PM ($B>5\,T$) states, respectively,
in agreement with earlier findings \cite{balicas_shubnikovhaas_2000,uji_two-dimensional_2001}.
The SdH oscillations also differ drastically in these two regimes.
In the PM state the oscillations originate from the classical $\alpha$ and
magnetic-breakdown (MB) $\beta$ orbits on the Fermi surface predicted by the tight-binding band
structure calculations \cite{kobayashi_new_1996}, see Fig.\,\ref{SdH}(b). In Fig.\,\ref{SdH}(c)
we show the fast Fourier transform (FFT) spectrum consisting of the fundamental harmonics,
$F_{\alpha} = 847$\,T and $F_{\beta} = 4230$\,T, and their combinations
\cite{balicas_shubnikovhaas_2000,uji_two-dimensional_2001,konoike_fermi_2005}.
By contrast, in the AF state only low-frequency oscillations, $F_{\delta}\approx 62$\,T, are observed,
indicating a Fermi surface reconstruction caused by the ordering \cite{konoike_fermi_2005}.

Moreover, even in the high-field PM state the SdH oscillations are affected by the $\pi$-$d$ coupling,
providing a basis for its quantitative evaluation.
In usual metals without magnetic interactions, the Zeeman splitting leads to a phase shift between
oscillations corresponding to spins parallel and antiparallel to the applied field.
This gives rise to a field-independent spin damping factor \cite{ShoenbergBook1984}, which in the
case of a strongly anisotropic layered metal has the form:
\begin{math}R_s=\cos\left( \frac{\pi}{2}gm^*_0/\cos\theta \right)\end{math}, where $g$ is the
Land\'{e} $g$-factor, $\theta$ is the angle between the magnetic field and the normal to conducting
layers (in our case the $b$-axis), and $m^*_0=m^*(\theta=0)$ the relevant cyclotron mass expressed in units
of the free electron mass.
However, when a constant internal field $B_J < 0$ is imposed on the conduction electrons, the spin factor becomes
field dependent \cite{ShoenbergBook1984,simp71}:
\begin{equation}
R_s=\cos\left[ \frac{\pi}{2}\frac{gm^*_0}{\cos\theta}\left(1- \frac{|B_J|}{B}\right) \right]\,,
\label{Rs}
\end{equation}
giving rise to periodic beats in the SdH amplitude with the frequency:
\begin{equation}
F_{\mathrm{b}}= \frac{gm^*_0|B_J|}{4\cos\theta}\,.
\label{Fbeat}
\end{equation}

In Fig.\,\ref{SdH}(c) the FFT peaks containing the MB frequency $F_{\beta}$ are split. C\'{e}pas et al.
\cite{cepas_magnetic-field-induced_2002} were the first to suggest that the splitting
$\Delta F_{\beta} \approx 100$\,T comes
from beating caused by the field-dependent spin factor in Eq.\,(\ref{Rs}). Indeed, the beat frequency,
$F_{\mathrm{b}} = \Delta F_{\beta}/2$, yields a reasonable value of the exchange field, $|B_J| = 12.5$\,T,
if we substitute $g = 2.0$, $m^*_{0,\beta} =8.0$ \cite{balicas_shubnikovhaas_2000} and
$\cos \theta =1$ (as $\mathbf{B} \perp$ layers). The problem, however, is that while the $\beta$ peak in
the SdH spectrum reproducibly shows splitting \cite{balicas_shubnikovhaas_2000,uji_two-dimensional_2001,konoike_fermi_2005},
no splitting has been reported for the $\alpha$-oscillations. In our measurements no sign of modulation of
the $\alpha$-oscillation amplitude has been found in the field window 5.5 to 14\,T.
This can bee seen from Fig.\,\ref{SdH}(d) where the amplitude $A_{\alpha}$ is plotted against inverse field,
revealing a monotonic exponential dependence. Fitting it with the standard formula for the field-dependent
oscillation amplitude yields the Dingle temperature (a quantity characterizing the scatttering
rate, $k_BT_D=\hbar/2\pi\tau$ \cite{ShoenbergBook1984}) $T_D = 0.25 \pm 0.3$\,K.
According to Eq.\,(\ref{Fbeat}), the absence of a minimum in the oscillation amplitude in the interval
between 5.5 and 14\,T would correspond to the exchange field being smaller than 0.9\,T (the cyclotron
mass $m^*_{0,\alpha} =5$ \cite{balicas_shubnikovhaas_2000,uji_two-dimensional_2001} was used for the estimation).
This controversy calls for a more thorough investigation of the effect of magnetic interactions on the SdH
oscillations.

\begin{figure}[tb]
	\centering
		\includegraphics[width=0.45\textwidth]{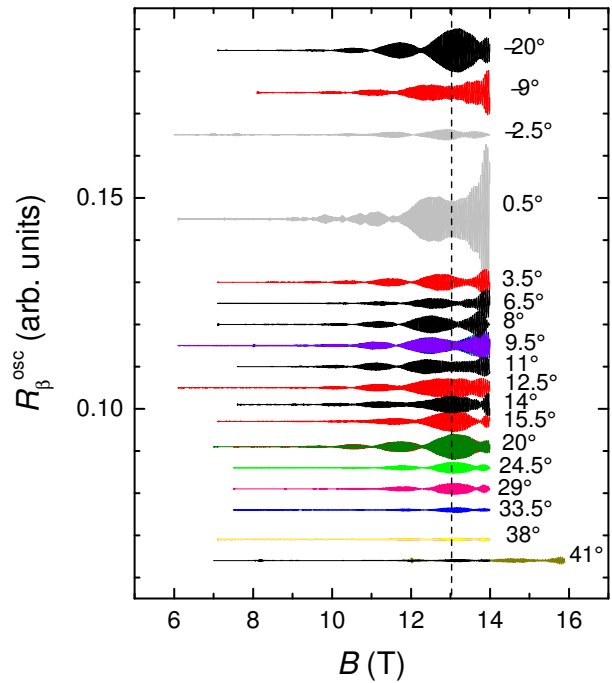}
	\caption{The $\beta$-oscillatory component of the magnetoresistance measured at $T=0.45$\,K at different
polar angles $\theta$. The curves are vertically shifted for clarity. The vertical line indicates the field,
at which the high-angle oscillations show a maximum: $B = -B_J$, according to Eq.\,(\ref{Rs})}
	\label{beta}
\end{figure}
To this end we have carried out experiments at different orientations of magnetic field.
Figure \ref{beta} shows the $\beta$-oscillating component of magnetoresistance measured at different polar
angles $\theta$. One clearly sees beats with the maxima and minima of the oscillation amplitude changing with
$\theta$. Considering the beats as coming from the field-dependent spin factor, the first, obvious
feature following from Eq.\,(\ref{Rs}) is that for $B=|B_J|$ the oscillation amplitude must always be at maximum,
independently of the angle. For $\theta \geq 20^{\circ}$ the behaviour is consistent with Eq.\,(\ref{Rs}), showing
the $\theta$-independent maximum position at $B\approx 13$\,T. However, for small tilt angles,
$-20^{\circ} < \theta < 20^{\circ}$,
the beats become more complex. Not only some of the beat nodes become much less pronounced but also their positions
shift in an apparently irregular manner. For example, at $\theta \approx \pm 9^{\circ}$ a minimum in the amplitude is
found around 13\,T, the field where the main ($\theta$-independent) maximum should be expected.

Using the condition for zeros of the spin factor $R_s$ determined by Eq.\,(\ref{Rs}), one can fit the positions of
the amplitude minima/nodes with the exchange field and $g$-factor as fitting parameters.
\begin{figure}[tb]
	\centering
		\includegraphics[width=0.45\textwidth]{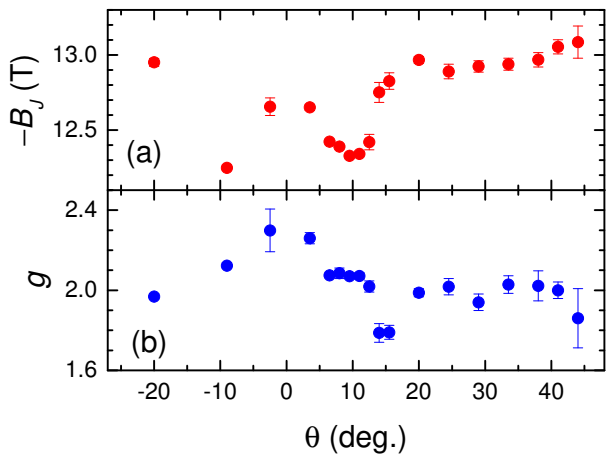}
	\caption{The $\pi$-$d$ exchange field and $g$-factor obtained from fitting the positions of the SdH amplitude
zeros/minima using Eq.\,(\ref{Rs})}
	\label{BJ&g}
\end{figure}
The results of such fitting are shown in Fig.\,\ref{BJ&g}. In line with what was qualitatively described above,
at $|\theta|\geq 20^{\circ}$ the obtained values, $B_J = -13.0 \pm 0.1$\,T and $g = 2.0 \pm 0.05$, are very reasonable
and consistent with the estimation based on the position of the re-entrant superconductivity region of the phase diagram,
see Sec. \ref{SC}. At lower tilt angles both fitting parameters strongly deviate from these values. At present we do not
have a satisfactory explanation of this anomalous behaviour. One might appeal to other possible mechanisms of beating of
the SdH oscillations have to be taken in consideration, such as weak warping of the Fermi surface in the interlayer
direction \cite{kart04} or mosaicity of the crystal.

\begin{figure}[tb]
	\centering
		\includegraphics[width=0.45\textwidth]{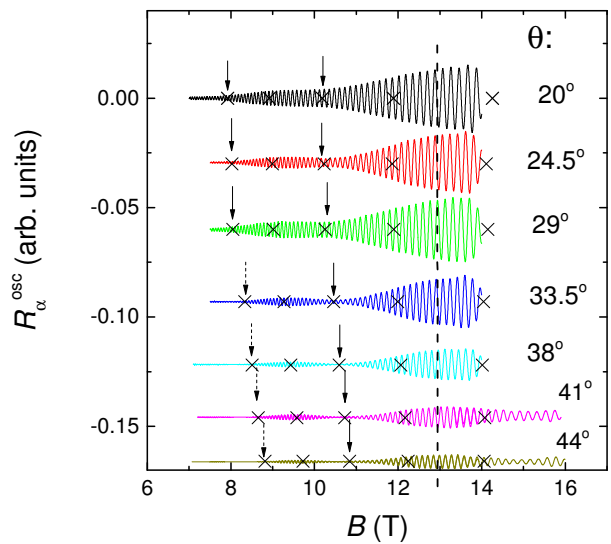}
	\caption{The $\alpha$-oscillatory component of the magnetoresistance measured at $T=0.45$\,K at different
polar angles $\theta$. The curves are vertically shifted for clarity. Crosses indicate the positions of the
amplitude minima expected from the analysis of the $\beta$-oscillations; arrows point to the actual minima
positions}
	\label{alpha}
\end{figure}

As noted above, the $\alpha$-oscillations corresponding to the smaller, classical orbit in Fig.\,\ref{SdH}(b) should
also be modulated by the field-dependent spin factor. However, we have found no modulation at tilt angles
below 15-20$^{\circ}$, i.e. in the range where the beat behaviour of the $\beta$-oscillations is also anomalous.
In Fig.\,\ref{alpha} we show the evolution of the $\alpha$-oscillating component at $\theta \geq 20^{\circ}$.
The weak modulation observed at 20$^{\circ}$ develops gradually into full beating with periodic nodes. However, the
node positions are still inconsistent with those found on the $\beta$-oscillations. As shown in Fig.\,\ref{alpha},
apparently only every second expected node shows up in the experiment. According to Eq.\,(\ref{Rs}), this would imply
that the $g$-factor on the $\alpha$ orbit is only one half of that on the $\beta$ orbit. This would be highly surprising,
especially since the latter does incorporate the $\alpha$ orbit as a part.
Thus, further angle-resolved studies are needed to clarify the situation. Perhaps low-$T$ ESR measurements will give a clue
to the puzzle.

\section{Summary}
The localized $d$-electron spins of Fe$^{3+}$ ions, responsible for magnetic properties, and itinerant electrons
originating from the $\pi$ orbitals of BETS molecules are spatially separated in the layered organic metal
$\kappa$-(BETS)$_2$\-FeBr$_4$. The exchange interaction between the two systems is weak, $J_{\pi d} \leq 1$\,meV.
Nevertheless this interaction has strong impact on the electronic properties.

The role of $\pi$-$d$ interaction is very clearly manifested in the high-field re-entrance superconductivity
\cite{fujiwara_indication_2002,konoike_magnetic-field-induced_2004}, but also in the low-field superconductivity,
which turns out to be "protected" by the AF ordering and disappear simultaneously with the latter at low
temperatures. IN the normal state, the $\pi$-$d$ coupling is reflected in the angle-dependent beating behaviour
of magnetic quantum oscillations. At sufficiently strongly tilted field, $|\theta| \geq 20^{\circ}$, the
beats of the magnetic-breakdown $\beta$-oscillations can be quantitatively accounted for by including the exchange
field $B_J = -13$\,T in the spin damping factor $R_s$ for the oscillation amplitude. However, at lower tilt angles
the beat positions are surprisingly shifted, resulting in an apparently strong changes in both $B_J$ and the $g$-factor.
This anomaly as well as the unexpected difference between the beats of the $\beta$- and $\alpha$-oscillations require
further investigation.

Finally, the conduction system in its turn seems to influence the magnetic ordering of the $d$-electron spins through
the indirect RKKY exchange. This is reflected in the change of the shape of the phase diagram and appearance of the
spin-flopped AF phase under pressure. It would be nice to perform magnetic measurements under pressure in order to
determine the exact nature of the high-field AF phase.

\section{Acknowledgement}

The work was supported by the German Research Foundation (DFG) via the grant KA 1652/4-1.




%


\end{document}